**Social Media Information Sharing for Natural Disaster Response**

Zhijie Sasha Dong[1], Lingyu Meng[1*], Lauren Christenson[1], Lawrence Fulton[2]

Ingram School of Engineering[1]
School of Health Administration[2]
Texas State University
San Marcos, Texas, 78666 USA
Corresponding author E-mail: l_m523@txstate.edu

**Abstract**

Social media has become an essential channel for posting disaster-related information, which provide governments and relief agencies real-time data for better disaster management. However, research in this field has not received sufficient attention and extracting useful information is still challenging. This paper aims to improve disaster relief efficiency via mining and analyzing social media data like public attitudes towards disaster response and public demands for targeted relief supplies during different types of disasters. We focus on different natural disasters based on properties such as types, durations, and damages, which contains a total of 41,993 tweets. In this paper, public perception is assessed qualitatively by manually classified tweets, which contain information like the demand for targeted relief supplies, satisfactions of disaster response, and public fear. Public attitudes to natural disasters are studied via a quantitative analysis using eight machine learning models. To better provide decision-makers with the appropriate model, the comparison of machine learning models based on computational time and prediction accuracy is conducted. The change of public opinion during different natural disasters and the evolution of people's behavior of using social media for disaster relief in the face of the identical type of natural disasters as Twitter continues to evolve are studied. The results in this paper demonstrate the feasibility and validation of the proposed research approach and provide relief agencies with insights into better disaster management.

**Keywords:** sentiment analysis, disaster response, big data analytics, machine learning, social media, twitter

**1 Introduction**

Within the decade, social media platforms, such as Facebook, Twitter, Instagram, Tumblr, LinkedIn, Reddit, Snapchat, and WeChat, have expanded on a global scale. Through virtual networks and groups people share their interests, opinions, attitudes, and other information via instantaneous communication (Martí et al. 2018).The emergence of these social media has spawned the user-generated content platform, which not only provides us opportunities to develop a computational technique for social media mining but also finds an original way for social science research (Lindsay 2012). Big data created from social media like Twitter has made a prominent position in almost all industries and sectors from individuals to government stakeholders, nongovernment institutions, private businesses, volunteering organizations, and so on. Nowadays, there has been a spurt of interest in the role of social media data and sentiment analysis in disaster response, since several natural disasters have struck across the globe every year, causing large-scale suffering and economic losses to the public (Beigi et al. 2016). In 2017 for instance, when Hurricane Harvey hit the Gulf of Mexico, thousands of national guard troops, police officers, and rescue workers helped the victims through messages posted on social media platforms asking for disaster relief (Li et al. 2019). The systematic use of social media can be beneficial for emergency management by receiving victim requests for assistance, monitoring user activities, and updating the public situational awareness among others (Lindsay 2012).

Administers can propose corresponding strategies based on learning and analyzing sentimental data, including public interests, opinions, and attitudes. For instance, Ranasinghe and Halgamuge (2020) studied the user behavior on cryptocurrencies via using the Twitter sentiment analysis for better trading decisions. Shabaz and Garg (2020) clustered Yelp's sentimental data for estimating the quality of stores. Shibuya and Tanaka (2018) considered the different characteristics of online communication posted by citizens after a large-scale disaster. However, it is a formidable task for a human reader to find the sentimental data hidden in the updated continuously and massive quantity of social media information. Consequently, automated sentiment discovery and summarization systems are needed. Sentiment analysis, sometimes also called opinion mining, grows out of this need. It is a popular subdiscipline of the broader field of Natural Language Processing (NLP), which is concerned with the classification of texts based on the expressed opinions or sentiments of the authors regarding a particular topic (Pang and Lee 2008). Advantages such as saving capital expenditures (e.g., time, money, and labor), tracking more people satisfactions, and identifying vital sentimental triggers making sentiment analysis one of the hottest topics for machine learning, especially in the field of NLP worldwide. However, machine learning models are usually domain-specific (Liu 2012; Wu et al. 2017). In other words, machine learning models do not work well on topics or text genres that are different. To solve the domain-specific feature, domain adaptation methods are needed (Alam et al. 2018; Glorot et al. 2011; Pan and Yang 2010). Domain adaptation uses the labeled data in the source domain and the unlabeled data in the target domain to train classifiers. As existing research rarely focused on disaster-related studies, one primary goal of this paper is to examine machine learning classifiers for natural disasters.

Social media such as Twitter have improved the crisis management due to the proliferation of the Internet and mobile technology (Lindsay 2012). Such resources offer a two-way communication platform for retrieving and disseminating disaster-related information (Haworth 2016). During a disaster, people can find useful information, provide first-hand reports of disasters, and seek help from governments via social media platforms (Nagy and Stamberger 2012; Neppalli 2017; Wang and Taylor 2018). First responders and relief agencies can identify the immediate needs of the affected population so that emergencies can be confirmed, and relief resources and actions can be allocated accordingly (Caragea et al. 2014; Kryvasheyeu et al. 2016). While social media is a promising resource for disaster management, it is still challenging to use these online platforms to identify information for decision making. The vast amount of data contains a wealth of unwanted information, which can be overwhelming and confusing for anyone trying to retrieve critical disaster-related information. These data characteristics also require high-performance extraction and analysis methods (Nazer et al. 2017; Olteanu et al 2015). Motivated by mining more significant information for disaster response, this paper not only conducts sentiment analysis, but also explores the weights of positive and negative sentimental data, the need for different relief supplies, and the time pattern of sending tweets regarding different types of natural disaster or the same type of naturasaster over time. Accordingly, the purpose of this research is to analyze people's sentimental characteristics during natural disasters by predicting opinioned texts so relief agencies, emergency managers, and government officers can provide targeted assistance based on public attitudes. Natural disasters act as a triggering factor for the spread of feeling and emotions through the internet, allowing for vast amounts of sentimental data to be collected. Social media data in disaster response can also aid in data analytics and information communication by detecting early warning messages, updating the disaster-related data, and monitoring the information sent by the public. Moreover, social media data contain many critical details, such as the essential relief supplies that the victims lack, the satisfaction that people feel, and the fear that the communities may have. Social media data analytics promises to be an emerging research approach to mitigate the devastation of natural disasters and can improve the effectiveness of disaster response, which is beneficial for humanitarian organizations, the government, as well as the general public.

The contributions of this paper are fourfold. First, we implement eight machine learning models in classifying public sentiment concerning disaster-related social media data and compare the efficiency of different machine learning based on both computational time and prediction accuracy. As the machine learning model has the domain-specific feature (Liu 2012; Wu et al. 2017), we make a comparison of these machine learning models via performance metrics such as confusion matrix and receiver operating characteristic (ROC) graph, which can help us analyze imbalanced datasets and improve classification accuracy. Second, we focus on analyzing social media data that contains public attitudes toward disaster response, especially essential supplies (e.g., food, housing, transportation, and medical supplies) that are in need and prioritized for victims. Compared with the existing research, which merely targets general positive or negative attitudes to natural disasters, a less generalized focus allows this research to have more practical implications. For instance, specific relief strategies and suggestions can be provided to both humanitarian organizations and victims according to our research focus. Third, we conduct this research from two perspectives for a comprehensive understanding of public opinion on disaster response. On the one hand, we explore public views on different types of natural disasters, including attitudes towards disaster response, demands for essential relief supplies, and the number of tweets posted online. On the other hand, this research concentrates on the change of public opinion on hurricanes, specifically because they are one of the most common and deadliest natural disasters in the United States. The thorough research can help the first responders to extend their horizons and serve the general public in a better way. Lastly, we create a natural disaster dataset with sentiment labels, which contains 41, 993 Twitter data points about natural disasters in the United States (Meng and Dong 2020). Based on the motivation of taking full advantage of the potential of social media data and saving capital expenditures, this dataset provides decision-makers like relief agencies, emergency managers, and government officers with an opportunity for future research.

The rest of this paper is organized as follows. Section 2 reviews relevant literature about social media and sentiment analysis in disaster response. Section 3 presents the description of the disaster-related dataset, including data collection. Section 4 demonstrates the quantitative analysis of the collected Twitter data. Section 5 explains the process of sentiment analysis, containing data preprocessing, training of machine learning models, evaluation of machine learning models, cross-validation and hyperparameter tuning, and prediction. Moreover, a set of case studies are discussed from the perspective of performance metrics like confusion matrix and ROC graph. Finally, conclusions are made in Section 6.

## 2. Literature review
### 2.1 Use of social media in disaster response
Social media has turned the web into a vast repository of data on many topics, which generates a potential source of information for science research (Batrinca and Treleaven 2014). In recent decades, social media has become a crucial information platform during natural disasters. Applications of social media in disaster response can be categorized into domains like disaster trajectory detection (Crooks et al. 2013; Earle et al. 2012), situational awareness (De et al. 2015; Power et al. 2014; Yin et al. 2012), information sharing (Li et al. 2018; Smith et al. 2017), and network performance evaluation (Kryvasheyeu et al. 2016; Wang and Taylor 2018). This paper focuses on situational awareness and information sharing, which are critical for improving disaster response. Situational awareness is identifying, processing, and comprehending essential elements of an incident or situation to provide useful insight into time and safety situations (Lindsay 2012; Ragini et al. 2018). It can assist first responders in assessing the number of damage and victims' needs. Information sharing is how people behave and share information on social media regarding various topics like natural disasters (Caragea et al. 2014; Kryvasheyeu et al. 2016). It can provide victims with a channel for communication (Nagy and Stamberger 2012; Neppalli 2017; Wang and Taylor 2018). Both situational awareness and information sharing are used for accelerating disaster response and alleviating devastations in natural disasters.

When Hurricane Katrina slammed the U.S. in 2005, people have not yet applied social media to natural disaster relief since Twitter did not exist, and Facebook was in its formative stage (Beigi et al. 2016). A few years later, a growing number of people started using Twitter, Facebook, Flicker, blogs, and YouTube to post their experience in the form of texts, photos, and videos during the Haiti earthquake on January 2010, resulting in 8 million U.S. dollars in donations to the Red Cross (Gao et al. 2011), and since then social media data has playing an increasing important role in disaster management. For example, Hughes et al. (2014) studied online public communications by police and fire services during the 2012 Hurricane Sandy and found it is important to consider online communication in future emergency management. Social media data has been used for warning individuals of emergency information and raising disaster relief funds for humanitarian organizations and governments in disasters including Hurricane Irene (Mandel et al. 2012), Genoa flooding (Buscaldi and Hernández-Farias 2015), and Ebola outbreak (Odlum and Yoon 2015). These studies successfully assisted humanitarian organizations in tracking, analyzing, and monitoring social media data related to disaster response (Calderon et al. 2014). Besides, Wang and Zhuang (2018) analyzed the rumor awareness and response behaviors of Twitter users during natural disasters. Lee et al. (2019) described the relationship among the precipitation deficit, the standardized precipitation index, the agricultural reservoir water storage deficit index, and news media data from agricultural drought-related news to evaluate the effectiveness of agricultural risk management using social media. Beigi et al. (2016) performed lexicon-based sentiment analysis of tweets posted on Twitter during the disastrous Hurricane Sandy and visualized online users' sentiments on a geographical map centered around the hurricane. Gabrielle et al. (2020) found social media may be a unique way to detect dietary patterns and help inform public health social media campaigns to advise people about stocking up on healthy, non-perishable foods ahead of natural disasters.

However, these studies generally focus on researching a specific natural disaster, rather than analyzing the change of people's opinions on different types of disasters. Moreover, they ignored the characteristic of public attitudes changing with time series when facing an identical natural disaster. This research attempt to fill these gaps.

### 2.2 Applications of sentiment analysis in disaster response
Sentiment analysis is the field of study that analyzes people's opinions, sentiments, and attitudes towards entities such as products, services, and individuals (Liu 2012; Pang and Lee 2014; Pak and Paroubek 2010). There are many studies on the impact of public opinion based on social media for various purposes, such as extracting political orientation (Ansari et al. 2020; Liu et al. 2016; Liu and Lei 2018), predicting markets' developing trends (Song et al. 2019; Si 2015; Vuong and Suzuki 2020), and utilizing artificial intelligence applications (Imran et al. 2014). In recent years, a great deal of researchers has conducted research about sentiment analysis. Some researchers focused on applications of sentiment analysis. Gadekallu et al. (2019) devised an aspect-oriented scheme that analyzes the textual reviews of a movie and assigned it a sentiment label on each aspect. Emotion vocabulary has been studied in various disciplines, such as psychology, linguistics, and computational linguistics by a research (Chen et al. 2018).

While others concentrated on developing sentiment analysis methods. For example, Zhao and Gui (2017) discussed the effects of text pre-processing method on sentiment classification performance in two types of classification tasks and summed up the classification performances of six pre-processing methods using two feature models and four classifiers on five Twitter datasets. A study (Agarwal et al. 2011) examined sentiment analysis on Twitter data and explored the use of a tree kernel to obviate the need for tedious feature engineering. Bing and Liang (2014) built a topic-based sentiment mixture model with topic-specific data in a semi-supervised training framework. Hatzivassiloglou et al. (2000) explored the benefits that some lexical features of adjectives offer for the prediction of a contextual sentence-level feature. Kennedy and Inkpen (2006) used a classifier model in simulating the effect of linguistic context. Sentiment analysis has been studied for more than two decades, with the accuracy and complexity of the methods increasing over time. One of the most common ways to analyze sentiment is the lexicon-based method (Hamilton et al. 2016; Kiritchenko et al. 2014), which utilizes a dictionary of pre-tagged words. To be more specific, each word in a text is compared against the dictionary with a polarity score, and its polarity value is added to the total polarity score of the text. Generally, if the total polarity score of a text is positive, then that text is classified as positive; otherwise, it is classified as negative (Annett 2008). Although naïve in nature, many variants of this lexicon-based method have been reported to perform better than chance.

With machine learning and big data analytics becoming popular, researchers continue to update research methods for better performance. The machine learning based sentiment analysis method has been developed (Tang et al. 2014; To et al. 2017). Since in general, supervised machine learning techniques perform better than the unsupervised machine learning techniques (Pan and Yang, 2010), this research utilized supervised machine learning models to accomplish classification tasks. To be more specific, a series of feature vectors are chosen, and a collection of labeled data is provided for training a classifier, which can then be applied to an unlabeled text. Bai and Yu (2016) proposed a structured machine learning method to detect potential incidents implicated by victims' negative emotions in the post-disaster situation. Ragini et al. (2018) proposed a methodology to visualize and analyze the sentiments of people affected by the disaster. Wang and Taylor (2020) developed a data-driven understanding of disaster dynamics in urban areas via Twitter data. Previous studies indicated the Naive Bayes algorithm and Support Vector Machines (SVM) algorithm are the most commonly employed classification techniques. The reported classification accuracy ranges between 63% and 82%, but these results are dependent upon the selected features.

Because sentiment analysis is usually domain-specific, our research compares the performance of different supervised machine learning models based on the characteristics of natural disaster dataset. Also, we will analyze the cause of this phenomenon.

## 3. Research methodology
The framework and approach used is this paper is schematically depicted in Fig. 1, which mainly contains four parts:identifying natural disasters, collecting raw data, preprocessing and analyzing disaster-relevant data qualitatively, implementing sentiment analysis via machine learning models. In this manner, this framework seeks to mine important information of social media data and explore how to take advantage of social media data for crisis management and disaster response.

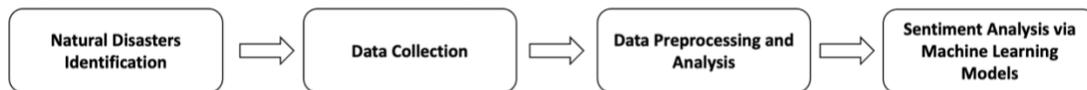

**Fig. 1** The framework of this study

### 3.1 Selection of natural disasters
Since the U.S. is frequently hit by various natural disasters, this paper studies the most destructive types of natural disasters over a decade based on Wikipedia statistics. We first select five different types of natural disasters that include a Tornado that occurred in April 2011, a series of floods that happened in September 2013, a Blizzard in January 2016, Hurricane Harvey in August 2017, and a set of wildfires which occurred in August 2018. We then focus on a specific type of disaster – hurricane because it is one of the most common and costly disasters in the United States. To be more precise, we selected Hurricane Sandy (October 2012), Hurricane Matthew (September 2016), Hurricane

Harvey (August 2017), Hurricane Michael (October 2018), and Hurricane Dorian (August 2019). Details obtained from Wikipedia, such as duration of disasters, economic loss, and fatality, are listed in Table 1 as follows.

**Table 1** Details of natural disasters

| Disaster | Duration | Economic Loss ($ billion) | Fatality |
|---|---|---|---|
| Tornado | 04/25/2011 ~ 04/28/2011 | 11 | 324 |
| Hurricane Sandy | 10/22/2012 ~ 11/02/2012 | 68.7 | 233 |
| Floods | 09/09/2013 ~ 12/31/2013 | 1 | 8 |
| Blizzard | 01/22/2016 ~ 01/24/2016 | 3 | 55 |
| Hurricane Matthew | 09/28/2016 ~ 10/10/2016 | 16.4 | 603 |
| Hurricane Harvey | 08/17/2017 ~ 09/02/2017 | 125 | 107 |
| Wildfires | 08/06/2018 ~ 11/08/2018 | 3.5 | 103 |
| Hurricane Michael | 10/07/2018 ~ 10/16/2018 | 25.1 | 74 |
| Hurricane Dorian | 08/24/2019 ~ 09/10/2019 | 4.68 | 84 |

### 3.2 Data collection

In this research, Twitter was chosen as the primary sentiment analysis object, as Twitter is a popular microblog that has 140 million active users posting more than 400 million tweets every day. A total of 41,993 valid tweets have been collected for these natural disasters. TwitterScraper was used to retrieve the content of Twitter in this research. The keywords that are used to collect these data are combinations of natural disasters and essential needs. The data was collected in the specified time frame, which is extended by one week before and after the duration of each natural disaster. Examples of keywords and time frames used for data collection are listed in Table 2 as follows.

**Table 2** Examples of keywords used for data collection

| Disaster | Keyword | Time Frame |
|---|---|---|
| Tornado | tornado + housing/transportation/food/medical supplies | 04/18/2011 - 05/05/2011 |
| Hurricane Sandy | hurricane sandy + housing/transportation/food/medical supplies | 10/15/2012 - 11/09/2012 |
| Floods | floods + housing/transportation/food/medical supplies | 09/02/2013 - 01/07/2014 |
| Blizzard | blizzard + housing/transportation/food/medical supplies | 01/15/2016 - 01/31/2016 |
| Hurricane Matthew | hurricane matthew + housing/transportation/food/medical supplies | 09/21/2016 - 10/17/2016 |
| Hurricane Harvey | hurricane harvey + housing/transportation/food/medical supplies | 08/10/2017 - 09/09/2017 |
| Wildfires | wildfires + housing/transportation/food/medical supplies | 07/31/2018 - 11/15/2018 |
| Hurricane Michael | hurricane michael + housing/transportation/food/medical supplies | 09/30/2018 - 10/23/2018 |
| Hurricane Dorian | hurricane dorian + housing/transportation/food/medical supplies | 08/17/2019 - 09/17/2019 |

### 3.3 Qualitative analysis

After data collection, this paper redefines the categorization of positive and negative sentiments based on the performance of government and relief agencies' disasters response. Based on the manual classification of sentimental tweets, a qualitative analysis is presented to demonstrate that analyzing social media data is useful for better disaster management.

### 3.4 Sentiment analysis

A quantitative analysis is conducted to provide decision-makers with an automatic tool for studying disaster-relevant social media data and solving the research problem. To be more specific, this paper implements eight machine learning models in classifying public sentiment concerning disaster-relevant social media data and comparing the efficiency of different machine learning based on both computational time and prediction accuracy. Results of this research allow emergency response managers to timely identify victims' demand and appropriately distribute relief supplies, which improves disaster relief efficiency.

## 4 A qualitative analysis of sentimental tweets

## 4.1 Determination of sentimental data

The determination of positive and negative sentiment will widely differ based on the subject and perspective. For the research problem in this paper, we only study positive and negative attitudes. We define the categorization of positive and negative sentiments based on the performance of government and relief agencies' disasters response. To be more specific, tweets containing information that governments or relief agencies make a positive impact on the disaster response are considered positive sentiments, otherwise are considered as negative sentiments. Our categorization can be more meaningful than just calibrating positive and negative feelings about the natural disaster itself. Table 3 presents some examples of sentimental labels. The positive attitude is assigned the label of 0, while the negative attitude is assigned the label of 1.

**Table 3** Examples of sentimental labels

| Label | Tweet |
| --- | --- |
| Positive | (1) "I am so happy that the Red Cross offers shelters for us" <br> (2) "Uber offers free rides to tornado victims staying in shelters" |
| Negative | (1) "Stores are empty like no food anywhere" <br> (2) "Why the governor has not given any evacuation instructions?" |

## 4.2 Analysis for different types of natural disasters

First, we have collected a total of 23,237 tweets for different types of natural disasters, of which Hurricane Harvey in 2017 has the most substantial proportion (32%), followed by Wildfires in 2018 (20%), Blizzard in 2016 (17%), Floods in 2013 (16%) and Tornado in 2011 (15%). Although the number of positive or negative sentiment in a disaster is affected by lots of factors, such as disaster severity, disaster type, and public vulnerability, we find that the proportion of negative tweets is higher than positive's regarding different natural disasters. Besides, if governments and rescue agencies issue warnings in advance, efficiently extricating people stranded by a disaster or providing sufficient essential supplies for victims, more positive sentiments would be collected on social media platforms. However, the proportion of negative sentiments outweighed the proportion of positive sentiments in all the natural disasters we studied, which indicates that the government and disaster relief agencies need to improve their disaster management and policies in general.

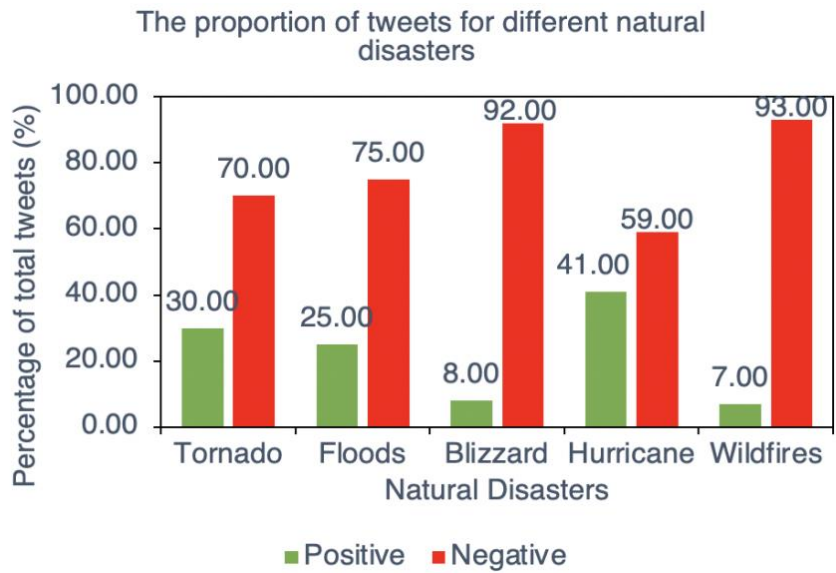

**Fig. 2** The proportion of tweets in different natural disasters

Second, we compare the number of tweets regarding different essential needs to explore public opinion further. We find that essential needs, which the general care most about, change in various natural disasters. For example, Fig. 3 illustrate that people conveyed their most concerns on housing, transportation, and food when the Tornado, Floods, and Hurricane Harvey happened, respectively. Therefore, humanitarian organizations should prepare appropriate rescue plans and essential relief supplies for these different situations. Meanwhile, we find transportation is the most emergent need in these relief supplies, which indicates that individuals need evacuation from the area that contains an ongoing threat to lives and property. Therefore, transportation sectors in the United States should pay more attention to traffic planning during emergency relief.

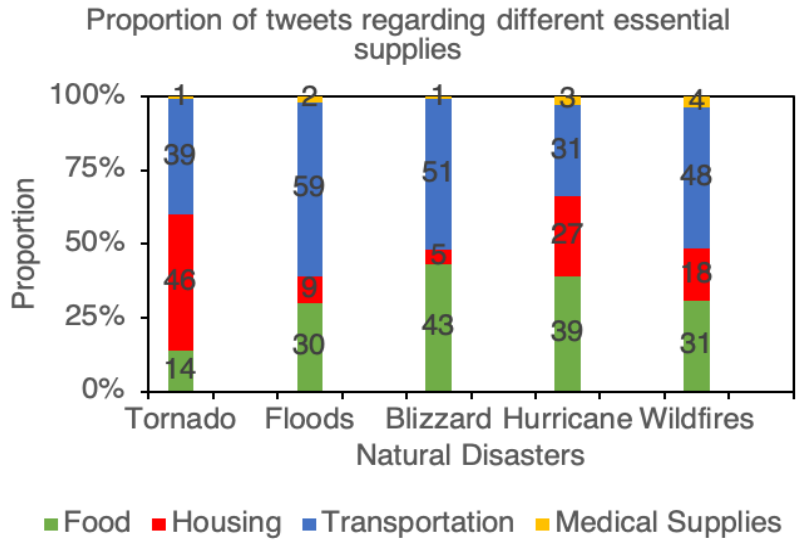

**Fig. 3** The proportion of tweets regarding different essential supplies

Third, we conduct an in-depth analysis of public attitudes about different essential needs. We find that although the public has negative attitudes towards all these necessary relief supplies, the share of negative views is different. For example, the public has the largest share of negative attitudes on transportation for a 2011 Tornado, medical supplies for Hurricane Matthew in 2016, and food for Hurricane Harvey in 2017. It can be seen that people have disparate levels of essential needs when natural disasters happen, so relief organizations should provide corresponding plans to natural disasters for a reasonable and efficient action of disaster response. Fig. 4 (a) – (e) present the details helping us provide targeted assistance for victims.

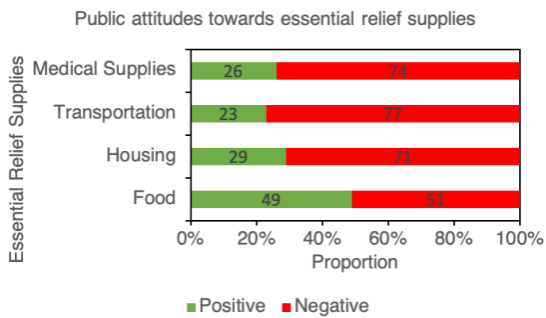

(a) 2011 Tornado

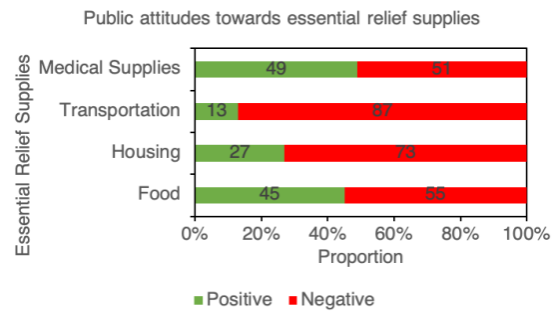

(b) 2013 Floods

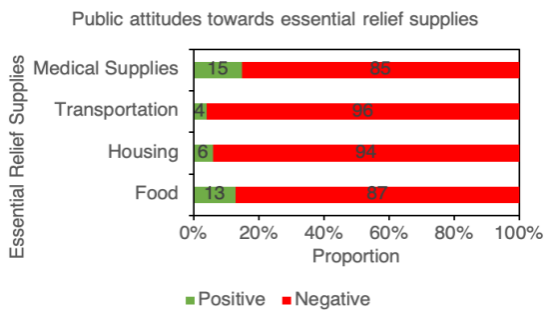

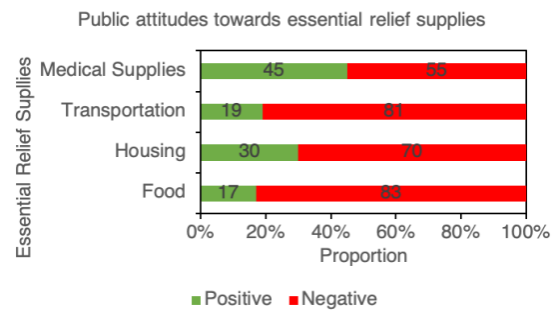

(c) 2016 Blizzard  (d) 2017 Hurricane

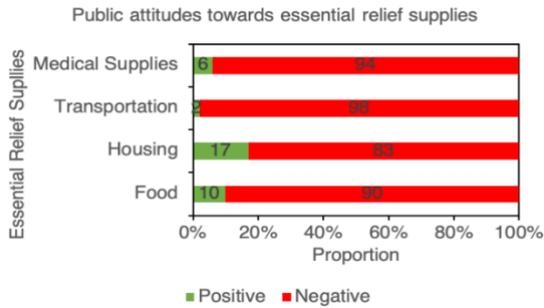

(e) 2018 Wildfires

**Fig. 4** Public attitudes towards essential needs

Lastly, we focus on studying the number of people and timing in which they publish their sentimental tweets on Twitter. To ensure data integrity, we not only examined changes in the number of tweets sent by people during a natural disaster (the red line in the figure below) but also focused on changes in the number of tweets before and after the disaster (the blue line in the figure below). Fig. 5 (a) – (e) depict that people have various trends for posting their tweets. For example, the public is likely to post tweets during the natural disasters such as the Tornado in 2011 and the Blizzard in 2016. However, people are intended to share their opinions after the period of the disaster for Wildfires in 2018. We assume this phenomenon is related to the attributes of different natural disasters. The public can have more time for disaster preparedness if natural disasters can be forecast (e.g., tornado, hurricane, and blizzard), and they will post their attitudes during the duration of disasters. In contrast, when people facing unpredictable hazards (e.g., wildfires and floods), they have the most concern about evacuating the hazardous area, and they are intended to convey their opinions later on.

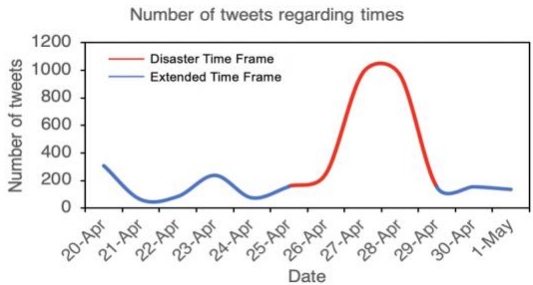 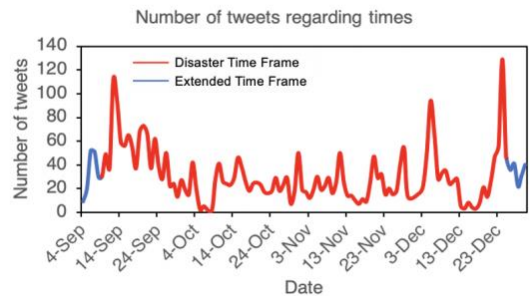

(a) 2011 Tornado                    (b) 2013 Floods

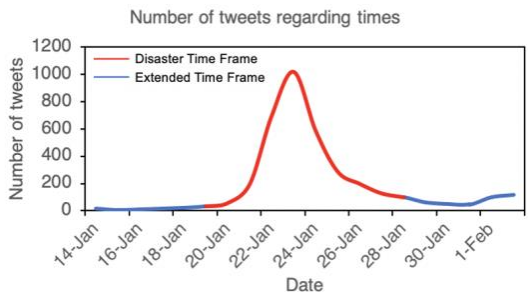 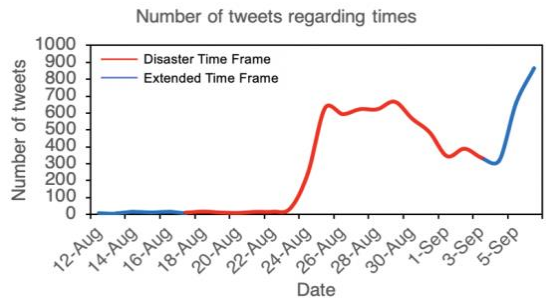

(c) 2016 Blizzard                   (d) 2017 Hurricane

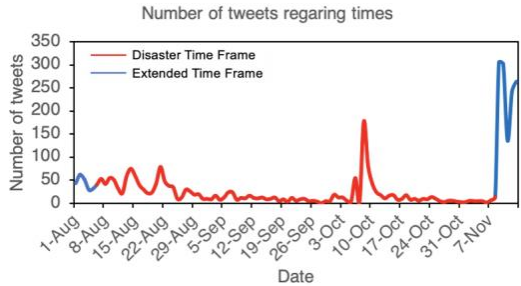

(e) 2018 Wildfires

**Fig. 5** The number of tweets in different natural disasters

**4.3 Analysis for the identical type of disaster**
In this section, we will study the same type of natural disaster that focuses on a time series analysis, and they are five hurricanes occurred in the United States. This research has collected a total of 26,579 tweets for these hurricanes, among which the Hurricane Harvey has the most substantial proportion (29%), followed by Hurricane Dorian (27%), Hurricane Matthew (20%), Hurricane Michael (16%) and Hurricane Sandy (8%). The proportion of positive and negative tweets obtained from Twitter are shown in Fig. 6. We find that the percentage of negative sentiments is higher than the rate of positive sentiments in the same type of natural disaster. (It is also consistent with previous results from studies of different kinds of natural disasters). Thus, it is possible to demonstrate that the conclusions we previously reached are earlier are vailed - the government and disaster relief agencies need to improve their disaster management and policies.

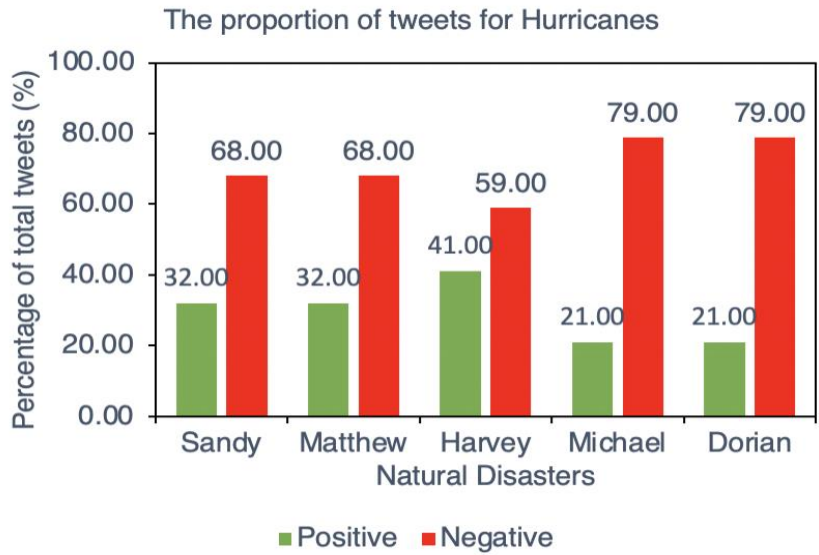

**Fig. 6** The proportion of tweets for Hurricanes

Fig. 6 illustrates that although we study the same type of natural disaster (i.e., hurricane), people also have different essential needs. For example, at the time of Hurricane Sandy, people cared more about housing. But when Hurricane Matthew happened, transportation was the primary concern. The supply relief that people were most concerned about is food at the time of Hurricane Michael. Fig. 7 shows that even when facing the same type of natural disaster, the essential needs of people can be different.

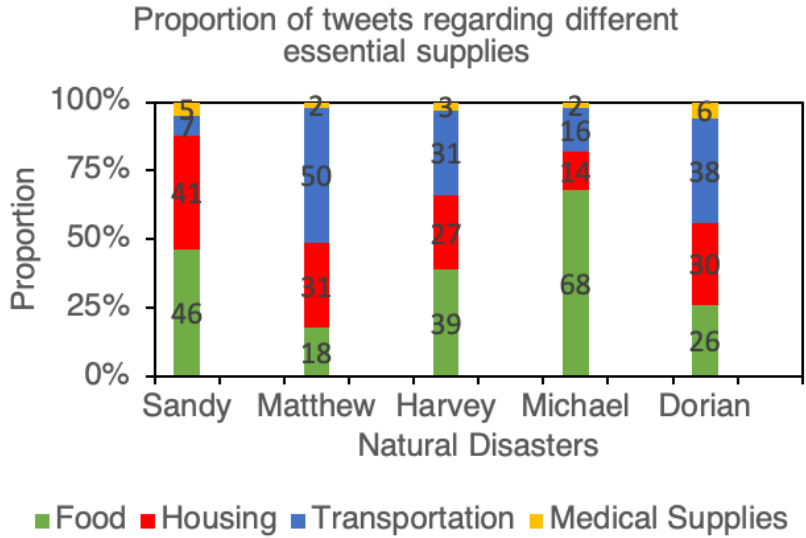

**Fig. 7** Proportion of tweets regarding different essential needs

Fig. 8 (a) – (e) presents that the proportion of positive attitudes towards essential relief supplies has increased over the years. Despite fluctuations in the percentage of positive sentiment, the overall trend is upwards. Except for 2016 Hurricane Matthew, sentiment towards medical supplies remained positive by more than 40%. People's sentiments about transportation have increased over time from a small proportion of positive sentiments to half of the positive sentiments. And people's positive sentiment about housing and food have not changed dramatically. It can seem that the effectiveness of disaster response has improved year by year. Thus, first responders should continue to draw on their experiences from past natural disasters to make better disaster management.

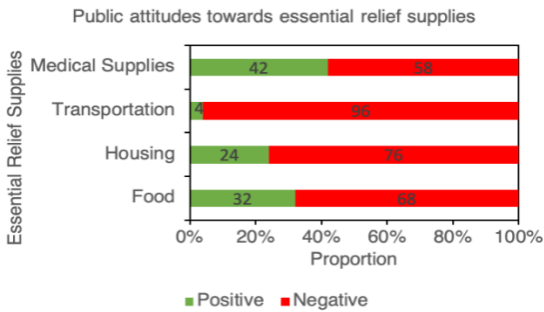

(a) 2012 Hurricane Sandy

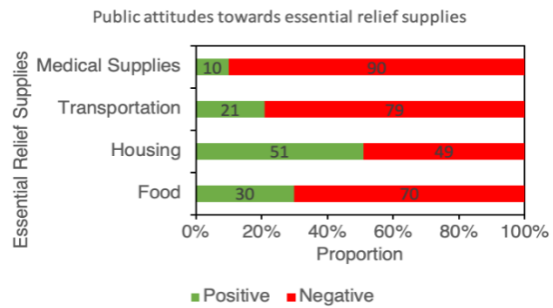

(b) 2016 Hurricane Matthew

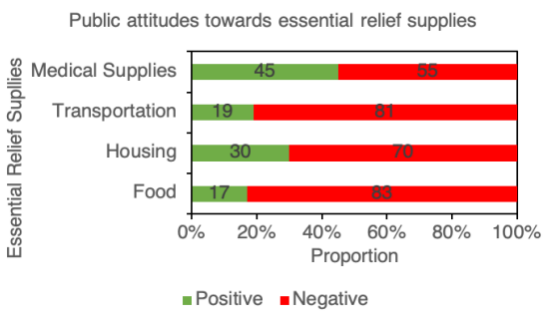

(c) 2017 Hurricane Harvey

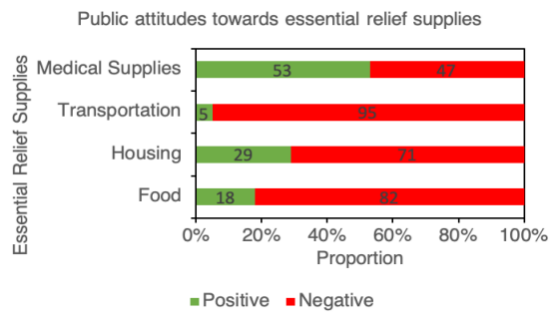

(d) 2018 Hurricane Michael

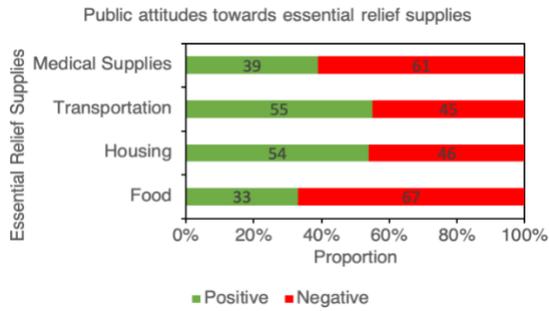

(e) 2019 Hurricane Dorian

**Fig. 8** Public attitudes towards essential relief supplies

Fig. 9 demonstrates that people are intended to post their attitudes at different times. As we can see from Fig. 9, people are used to posting tweets in the middle and latter stages of the disaster when facing a natural disaster like a hurricane. Since hurricane is a more common natural disaster in the U.S. and governments usually issue evacuation warning or release disaster relief policies in anticipation of its arrival, people have adequate supplies in the run-up to such natural disasters. Besides, people may be too busy to prepare for an evacuation to update their social media in the first place. Therefore, at this point, researchers can obtain public opinions after a hurricane happens. A real-time data analysis platform can be built to make it easier for the public to exchange information and promote disaster response to relief goods.

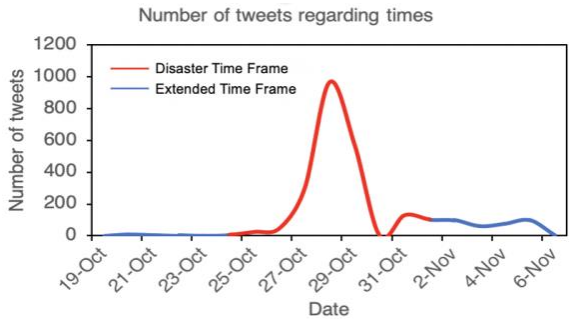

(a) 2012 Hurricane Sandy

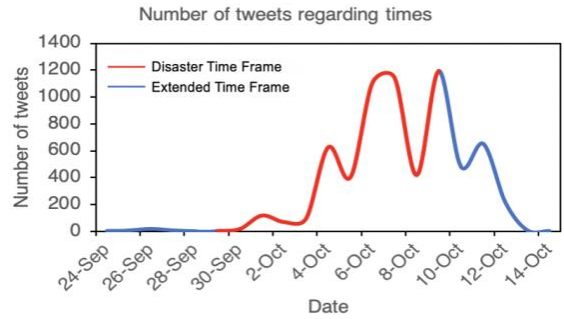

(b) 2016 Hurricane Matthew

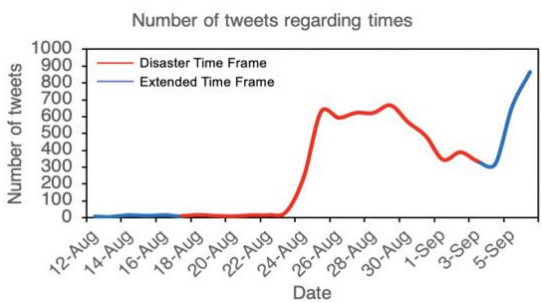

(c) 2017 Hurricane Harvey

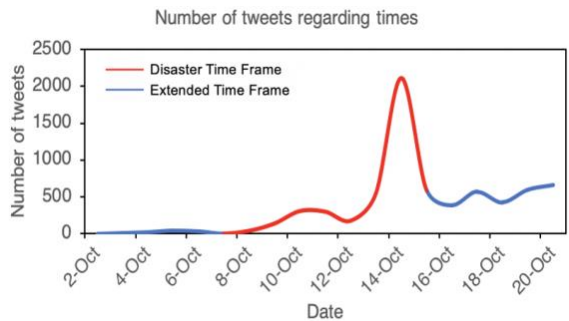

(d) 2018 Hurricane Michael

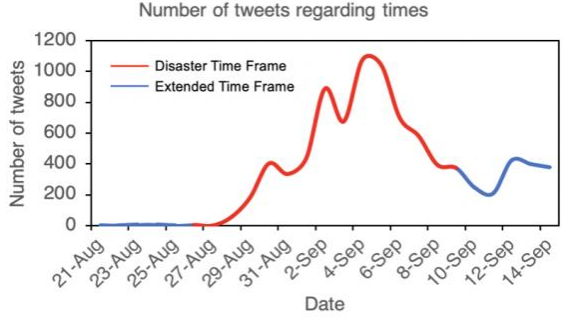

(e) 2019 Hurricane Dorian

**Fig. 9** The number of tweets in the identical natural disasters

## 5. A quantitative analysis using machine learning models

The quantitative analysis of the previous section is based on sentimental data that have been manually labeled. It took two members of our team nearly two months to complete the manual calibration of the 41,993 sentimental data. As this paper mentioned before, however, it is a formidable task for a human to find the sentimental data hidden in the updated continuously and massive quantity of social media information. Besides, disaster management is a time-sensitive process. It requires governments and disaster responders to analyze the feeling and need for victims quickly. Therefore, people need a machine learning approach to improve the efficiency of the analysis. In this section, machine learning models for sentiment analysis are explored for better disaster management.

The procedures for implementing machine learning models for sentiment analysis in this paper consist of five tasks: data collection, data preprocessing, learning, evaluation, and prediction. Data collection has been introduced in the previous section, and we will present the remaining tasks in this section. Fig. 10 shows the workflow for developing machine learning models for sentiment analysis.

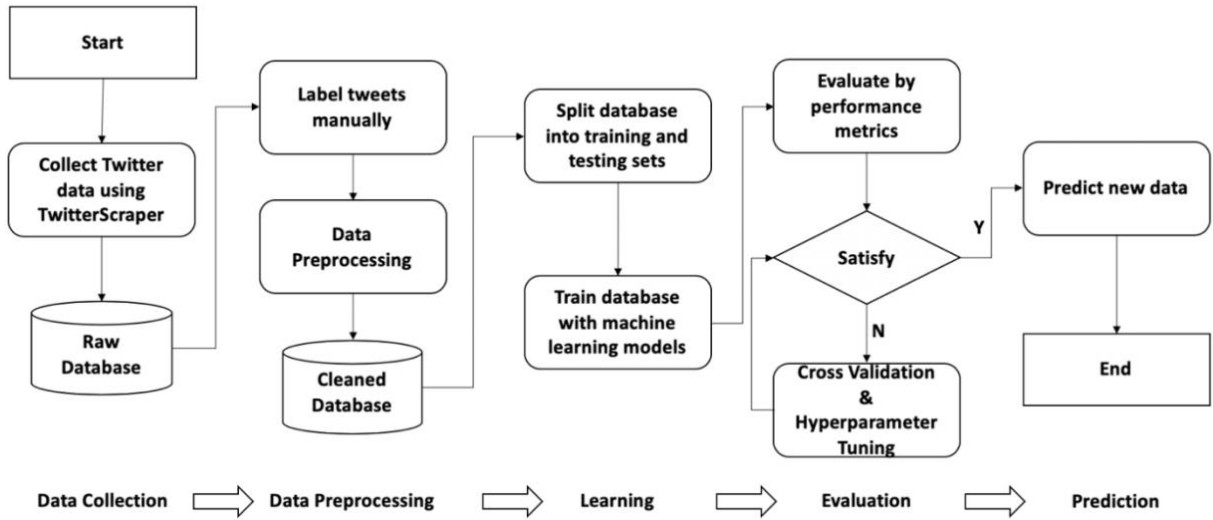

**Fig. 10** Workflow for implementing machine learning models for sentiment analysis

### 5.1 Data preprocessing
#### 5.1.1 Data cleaning
The quality of the dataset affects how well a machine learning model can learn significantly. Therefore, we must make sure to examine and preprocess a dataset before we feed it to a learning model. In general, the raw Twitter data contains HTML markup as well as punctuation and other special characters. For simplicity, this research removes all punctuation marks, weblinks, and useless information in the raw Twitter data. To accomplish this task, we will use Python's regular expression (regex) library. Then we need to split the raw tweet into individual words for further data preprocessing, such as modifying individual words to their root forms, converting the letters of each word to lowercases, and removing stop-words. Stop-words are those words that are extremely common in all sorts of texts and probably bear no useful information that can be used to distinguish between different classes of documents. Examples of stop-words are "is", "and", "has", and "like". The Natural Language Toolkit (NLTK) is a powerful package for python to implement cleaning, which is used in this research to accomplish the data preprocessing task.

### 5.1.2 Feature vector
We have to convert categorical data, such as text or words, into a numerical form before we can pass it on to a machine learning model. In this research, we use a useful technique called term frequency-inverse document frequency (tf-idf) that can be used to transform words into feature vectors. The tf-idf can be defined as a product of the term frequency and the inverse document frequency:

$$\text{tf} - \text{idf}(t, d) = tf(t, d) * idf(t, d) \tag{1}$$

where $tf(t, d)$ is the term frequency, which indicates the number of times that a term $t$ occurs in a document $d$. $idf(t, d)$ is the inverse document frequency and can be calculated as follows:

$$idf(t, d) = log \frac{n_d}{1 + df(d, t)} \tag{2}$$

where $n_d$ is the total number of documents, and $df(d, t)$ is the number of documents $d$ that contain the term $t$. Note that adding the constant 1 to the denominator is optimal and it serves the purpose of assigning a non-zero value to terms that occur in all training samples. The log is used to ensure that low document frequencies are not given too much weight.

### 5.2 Learning
Machine learning algorithms include supervised learning, unsupervised learning, and reinforcement learning. Supervised learning learns by using labelled data, which deals with regression (cluster continuous data) and classification (predict a discrete value) problems. This algorithm aims to calculate outcomes, including risk evaluation, forecast sales, and predict the trend. Unsupervised learning trains using unlabeled data without any guidance, which deals with association and clustering. Discovering underlying patterns is unsupervised learning's target. Applications of this algorithm include recommendations system and anomaly detection. Reinforcement learning is neither based on supervised learning nor unsupervised learning. It works on interacting with the environment and aiming at learning a series of actions. Self-driving cars, gaming, and healthcare system's development are supported by the reinforcement learning.

Supervised learning is used since all the machine learning models in this paper learns by discrete labelled data. Also, the purpose of this paper is to explore the possible role of social media in natural disaster relief by predicting users' positive and negative sentiments, which belongs to classification problem. To compare the performance of different machine learning models, we use eight common and popular models to classify the natural disasters datasets into positive and negative attitudes. These machine learning models include Logistic Regression (LR), Naïve Bayes (NB), Decision Tree (DT), Support Vector Machine (SVM), KNeighbor (KNN), Random Forests (RF), Adaboost, and Multiple Neutral Network (MNN). Fig 4 shows the advantages and disadvantages of the different models.

**Table 4** Advantages and disadvantages of different machine learning models

|    | Advantages | Disadvantages |
|----|------------|---------------|
| NB | (1) They are easy to implement and can scale with the dataset. | (1) Due to their sheer simplicity, NB models are often beaten by models properly trained and tuned using the previous algorithms listed. |
| LR | (1) Outputs have a nice probabilistic interpretation. (2) The algorithm can be regularized to avoid overfitting. (3) They can be updated easily with new data using stochastic gradient descent. | (1) LR tends to underperform when there are multiple or non-linear decision boundaries. (2) They are not flexible enough to naturally capture more complex relationships. |
| DT | (1) Requires little data preparation.<br>(2) The cost of using the tree (i.e., predicting data) is logarithmic in the number of data points used to train the tree.<br>(3) Able to handle both numerical and categorical data. | (1) Decision trees can be unstable because small variations in the data might result in a completely different tree being generated.<br>(2) The problem of learning an optimal decision tree is known to be NP-complete under several aspects of optimality and even for simple concepts. |
| SVM | (1) Effective in high dimensional spaces.<br>(2) Still effective in cases where number of dimensions is greater than the number of samples.<br>(3) Uses a subset of training points in the decision function (called support vectors), so it is also memory efficient. | (1) If the number of features is much greater than the number of samples, avoid over-fitting in choosing Kernel functions and regularization term is crucial.<br>(2) SVMs do not directly provide probability estimates, these are calculated using an expensive five-fold cross-validation. |
| KNN | (1) Robust to noisy training data.<br>(2) Effective if the training data is large | (1) Need to determine value of parameter K.<br>(2) Distance based learning is not clear which type of distance to used and which attribute to use to produce the best results.<br>(3) Computational cost is quite high. |
| RF | (1) As with regression, classification tree ensembles also perform very well in practice.<br>(2) They are robust to outliers, scalable, and able to naturally model non-linear decision boundaries thanks to their hierarchical structure. | (1) Unconstrained, individual trees are prone to overfitting, but this can be alleviated by ensemble methods. |
| AdaBoost | (1) As with regression, classification tree ensembles also perform very well in practice.<br>(2) They are robust to outliers, scalable, and able to naturally model non-linear decision boundaries thanks to their hierarchical structure. | (1) Unconstrained, individual trees are prone to overfitting, but this can be alleviated by ensemble methods. |
| MNN | (1) Capability to learn non-linear models.<br>(2) Capability to learn models in real-time | (1) MNN with hidden layers have a non-convex loss function where there exists more than one local minimum.<br>(2) MNN requires tuning a number of hyperparameters such as the number of hidden neurons, layers, and iterations.<br>(3) MNN is sensitive to feature scaling. |

For each disaster, we divide the entire dataset into training and testing categories, with the training set accounting for 30% of the total dataset and the testing set accounting for the remaining of the dataset. As shown in Fig. 11 below, each tweet has two features: X represents the text of the tweet, and Y represents the sentimental label of the tweet. In the learning task, we focus on these two features of the dataset and then we can input the values of X and Y into different models for training.

| Text (X) | Label (Y) |
|---|---|
| RT @ bedground: Please RT – Food needed for shelter in NC in area that was hit by tornado. | 1 |
| Although Hurricane Matthew storms towards Haiti and Jamaica, families are well prepared emergency supplies. | 0 |

**Fig. 11** Features of tweets

**5.3 Model evaluation**

This chapter uses a GridSearchCV object to find the optimal set of parameters for these training models using 5-fold stratified cross-validation. The GridSearchCV is a process of selecting the values for a model's parameters that maximize the accuracy of the model, and the cross-validation is a process of training a model using a set of data and testing it using a different set. We need to evaluate our models because it helps us better use our data, and it gives us much more information about our algorithm's performance. Last but not least, this procedure can effectively prevent the problem of underfitting and overfitting. For example, the logistic regression model contains parameters such as penalty, c-value, and the range of n-gram. We can hardly find the optimal parameter combination of the logistic regression model manually, while GridSearchCV allows us to fit a model including an arbitrary number of transformation steps and apply it to make classification about new data.

**5.4 Prediction**
The data is cleaned according to the data preprocessing task is given as input to the machine learning model, which has been optimized by model evaluation task. The prediction results of different machine learning models are evaluated by calculating Confusion Matrix and plotting Receiver Operating Characteristic (ROC) graphs.

**5.4.1 Confusion matrix**
The confusion matrix is a square matrix that reports the counts of the True Positive (TP), True Negative (TN), False Positive (FP), and False Negative (FN) predictions of a machine learning model, as shown in the Fig. 12.

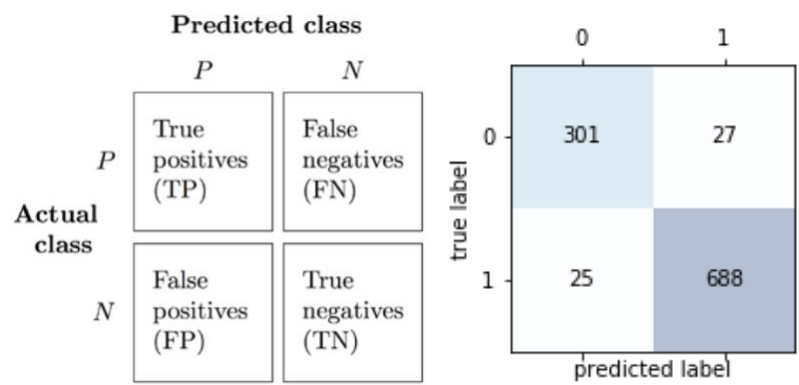

**Fig. 12** Confusion matrix

In a confusion matrix, these four main metrics for measuring the performance of a classification model are as follows. Accuracy (ACC) is the general information about how many samples are classified correctly. Precision (PRE) is a fraction of the classified text that are relevant. Recall (REC) is a fraction of the classified text that are retrieved. F1-score (F1) is the harmonic mean of PRE and REC.

Tables 5 - 6 show the results of the natural disaster dataset classification using different machine learning models. We find that the Naïve Bayes model has the best average prediction accuracy, followed by the Logistics Regression model, Random Forest model, AdaBoost model, SVM model, KNN model, MNN model, and Decision Tree model. It indicates that machine learning models, like linear and logistic models, have higher classification accuracy than complicated machine learning models like ensemble models and deep learning models regarding the text classification problem like sentiment analysis. This is most likely that complicated machine learning models are better at dealing with complex problems like image and audio processing. Meanwhile, the amount of data in this research is limited so that more rudimentary machine learning models may obtain a better classification result. Also, we find that several metrics can be used to measure a model and they can tell different stories about our machine learning models. Generally, accuracy is the most intuitive performance measure, which is a ratio of correctly predicted samples to the total observations. However, accuracy cannot tell all the features of a machine learning model, especially when the dataset is imbalanced. For example, in Table 5 we find that Naïve Bayes model has a high recall and precision value regarding both positive and negative reviews, which indicates the class is perfectly handled by the model. However, models like SVM, Adaboost, and MNN have either a low recall and high precision value or high recall and low precision value, which means the model cannot detect the class well but is highly trustable when it does or the class is well detected, but the model also includes other points in it. Thus, we introduce the Receiver Operating Characteristic (ROC) graphs for further analysis.

**Table 5** Accuracies of the classification

|  | Naïve Bayes | Logistic Regression | Decision Tree | SVM |
|---|---|---|---|---|
| Tornado | 0.94 | 0.73 | 0.66 | 0.75 |
| Hurricane Sandy | 0.97 | 0.85 | 0.78 | 0.81 |
| Floods | 0.97 | 0.73 | 0.65 | 0.78 |
| Blizzard | 0.98 | 0.89 | 0.78 | 0.90 |
| Matthew | 0.95 | 0.95 | 0.89 | 0.91 |
| Hurricane Harvey | 0.90 | 0.87 | 0.75 | 0.84 |
| Hurricane Michael | 0.98 | 0.91 | 0.80 | 0.88 |
| Wildfires | 0.99 | 0.93 | 0.87 | 0.92 |
| Hurricane Dorian | 0.95 | 0.93 | 0.91 | 0.93 |
| Average | 0.96 | 0.87 | 0.79 | 0.86 |
|  | KNN | Random Forest | AdaBoost | MNN |
| Tornado | 0.64 | 0.77 | 0.73 | 0.70 |
| Hurricane Sandy | 0.78 | 0.80 | 0.82 | 0.81 |
| Floods | 0.77 | 0.78 | 0.76 | 0.81 |
| Blizzard | 0.92 | 0.93 | 0.92 | 0.68 |
| Matthew | 0.87 | 0.93 | 0.95 | 0.83 |
| Hurricane Harvey | 0.68 | 0.85 | 0.82 | 0.84 |
| Hurricane Michael | 0.87 | 0.88 | 0.91 | 0.87 |
| Wildfires | 0.94 | 0.94 | 0.92 | 0.75 |
| Hurricane Dorian | 0.91 | 0.93 | 0.92 | 0.88 |
| Average | 0.82 | 0.87 | 0.86 | 0.80 |

**Table 6** Metrics of the classification

| Disaster | Label | Naïve Bayes | | | Logistics Regression | | | Decision Tree | | | SVM | | |
|---|---|---|---|---|---|---|---|---|---|---|---|---|---|
|  |  | P | R | F1 | P | R | F1 | P | R | F1 | P | R | F1 |
| Tornado | Positive | 1.00 | 0.81 | 0.90 | 0.55 | 0.53 | 0.54 | 0.44 | 0.55 | 0.49 | 0.59 | 0.48 | 0.53 |
|  | Negative | 0.93 | 1.00 | 0.96 | 0.81 | 0.82 | 0.81 | 0.79 | 0.70 | 0.74 | 0.80 | 0.86 | 0.83 |
| Hurricane Sandy | Positive | 1.00 | 0.91 | 0.95 | 0.76 | 0.75 | 0.76 | 0.63 | 0.71 | 0.67 | 0.88 | 0.47 | 0.62 |
|  | Negative | 0.96 | 1.00 | 0.98 | 0.88 | 0.89 | 0.89 | 0.86 | 0.81 | 0.83 | 0.80 | 0.97 | 0.88 |
| Floods | Positive | 1.00 | 0.87 | 0.93 | 0.45 | 0.54 | 0.49 | 0.36 | 0.51 | 0.42 | 0.58 | 0.43 | 0.49 |
|  | Negative | 0.96 | 1.00 | 0.98 | 0.84 | 0.79 | 0.81 | 0.81 | 0.70 | 0.75 | 0.83 | 0.90 | 0.86 |
| Blizzard | Positive | 1.00 | 0.81 | 0.89 | 0.38 | 0.48 | 0.42 | 0.18 | 0.44 | 0.26 | 0.41 | 0.44 | 0.42 |
|  | Negative | 0.98 | 1.00 | 0.99 | 0.95 | 0.93 | 0.94 | 0.94 | 0.82 | 0.87 | 0.95 | 0.94 | 0.94 |
| Hurricane Matthew | Positive | 1.00 | 0.84 | 0.91 | 0.94 | 0.94 | 0.94 | 0.78 | 0.89 | 0.83 | 0.85 | 0.88 | 0.87 |
|  | Negative | 0.93 | 1.00 | 0.96 | 0.95 | 0.95 | 0.95 | 0.95 | 0.88 | 0.91 | 0.95 | 0.93 | 0.94 |
| Hurricane Harvey | Positive | 1.00 | 0.75 | 0.86 | 0.84 | 0.85 | 0.84 | 0.68 | 0.72 | 0.70 | 0.76 | 0.88 | 0.82 |
|  | Negative | 0.85 | 1.00 | 0.92 | 0.89 | 0.89 | 0.89 | 0.80 | 0.77 | 0.78 | 0.91 | 0.81 | 0.86 |
| Hurricane Michael | Positive | 1.00 | 0.92 | 0.96 | 0.78 | 0.81 | 0.80 | 0.52 | 0.66 | 0.58 | 0.67 | 0.84 | 0.74 |
|  | Negative | 0.98 | 1.00 | 0.99 | 0.95 | 0.94 | 0.94 | 0.90 | 0.84 | 0.87 | 0.95 | 0.89 | 0.92 |
| Wildfires | Positive | 1.00 | 0.90 | 0.94 | 0.49 | 0.48 | 0.48 | 0.29 | 0.54 | 0.38 | 0.44 | 0.57 | 0.49 |
|  | Negative | 0.99 | 1.00 | 1.00 | 0.96 | 0.96 | 0.96 | 0.96 | 0.90 | 0.93 | 0.97 | 0.94 | 0.95 |
| Hurricane Dorian | Positive | 1.00 | 0.75 | 0.86 | 0.84 | 0.84 | 0.84 | 0.73 | 0.88 | 0.80 | 0.80 | 0.85 | 0.83 |
|  | Negative | 0.94 | 1.00 | 0.97 | 0.96 | 0.96 | 0.96 | 0.97 | 0.92 | 0.94 | 0.96 | 0.95 | 0.95 |
| Disaster | Label | KNN | | | Random Forest | | | AdaBoost | | | MNN | | |
|  |  | P | R | F1 | P | P | R | F1 | P | P | R | F1 | P |
| Tornado | Positive | 0.43 | 0.60 | 0.50 | 0.78 | 0.29 | 0.43 | 0.56 | 0.37 | 0.45 | 0.50 | 0.50 | 0.50 |
|  | Negative | 0.80 | 0.66 | 0.72 | 0.77 | 0.96 | 0.85 | 0.77 | 0.88 | 0.82 | 0.79 | 0.79 | 0.79 |
| Hurricane Sandy | Positive | 0.92 | 0.34 | 0.49 | 0.85 | 0.46 | 0.60 | 0.76 | 0.65 | 0.70 | 0.67 | 0.81 | 0.73 |
| | Negative | 0.76 | 0.99 | 0.86 | 0.79 | 0.96 | 0.87 | 0.85 | 0.91 | 0.88 | 0.90 | 0.82 | 0.86 |
| Floods | Positive | 0.92 | 0.07 | 0.13 | 0.66 | 0.21 | 0.32 | 0.51 | 0.28 | 0.36 | 0.44 | 0.52 | 0.47 |
| | Negative | 0.77 | 1.00 | 0.87 | 0.79 | 0.96 | 0.87 | 0.80 | 0.91 | 0.85 | 0.83 | 0.78 | 0.81 |
| Blizzard | Positive | 0.62 | 0.24 | 0.35 | 0.68 | 0.24 | 0.36 | 0.50 | 0.26 | 0.34 | 0.15 | 0.58 | 0.23 |
| | Negative | 0.93 | 0.99 | 0.96 | 0.93 | 0.99 | 0.96 | 0.93 | 0.98 | 0.95 | 0.95 | 0.68 | 0.79 |
| Hurricane Matthew | Positive | 0.95 | 0.62 | 0.75 | 0.97 | 0.80 | 0.88 | 0.95 | 0.89 | 0.92 | 0.68 | 0.91 | 0.77 |
| | Negative | 0.85 | 0.98 | 0.91 | 0.92 | 0.99 | 0.95 | 0.95 | 0.98 | 0.96 | 0.95 | 0.80 | 0.87 |
| Hurricane Harvey | Positive | 0.62 | 0.54 | 0.58 | 0.83 | 0.80 | 0.82 | 0.81 | 0.72 | 0.76 | 0.81 | 0.80 | 0.80 |
| | Negative | 0.71 | 0.78 | 0.74 | 0.87 | 0.89 | 0.88 | 0.82 | 0.88 | 0.85 | 0.86 | 0.87 | 0.87 |
| Hurricane Michael | Positive | 1.00 | 0.40 | 0.58 | 0.69 | 0.74 | 0.72 | 0.84 | 0.71 | 0.77 | 0.68 | 0.74 | 0.71 |
| | Negative | 0.86 | 1.00 | 0.93 | 0.93 | 0.91 | 0.92 | 0.93 | 0.96 | 0.94 | 0.93 | 0.91 | 0.92 |
| Wildfires | Positive | 0.88 | 0.22 | 0.36 | 0.64 | 0.27 | 0.38 | 0.44 | 0.34 | 0.39 | 0.18 | 0.73 | 0.30 |
| | Negative | 0.94 | 1.00 | 0.97 | 0.95 | 0.99 | 0.97 | 0.95 | 0.97 | 0.96 | 0.97 | 0.75 | 0.84 |
| Hurricane Dorian | Positive | 0.98 | 0.59 | 0.74 | 0.97 | 0.70 | 0.81 | 0.85 | 0.75 | 0.80 | 0.64 | 0.91 | 0.75 |
| | Negative | 0.90 | 1.00 | 0.95 | 0.93 | 0.99 | 0.96 | 0.94 | 0.97 | 0.95 | 0.91 | 0.87 | 0.92 |

### 5.4.2 Receiver operating characteristic (ROC) graphs

Receiver Operating Characteristic (ROC) graphs are a useful tool to select models for classification based on their performance with respect to the FPR (False Positive Rate) and TPR (True Positive Rate), which are computed by shifting the decision threshold of the classifier. The diagonal of a ROC graph can be interpreted as random guessing, and classification models that fall below the diagonal are considered as worse than random guessing. A perfect classifier would fall into the top left corner of the graph with a TPR of 1 and FPR of 0. Based on the ROC curve, we can then compute the so-called ROC Area Under the Curve (ROC-AUC) to characterize the performance of a classification model. The True Positive Rate (TPR) and False Positive Rate (FPR) are performance metrics that are especially useful for imbalanced class problems. Here we use 2016 Hurricane Matthew as an example, Fig. 13 (a) – (h) shows the ROC curves for different machine learning models. Details of all the ROC-AUC values are shown in Table 7.

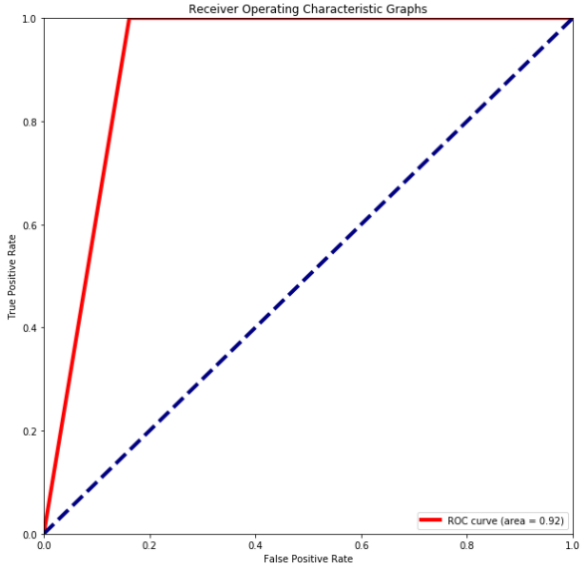

(a) Naïve Bayes

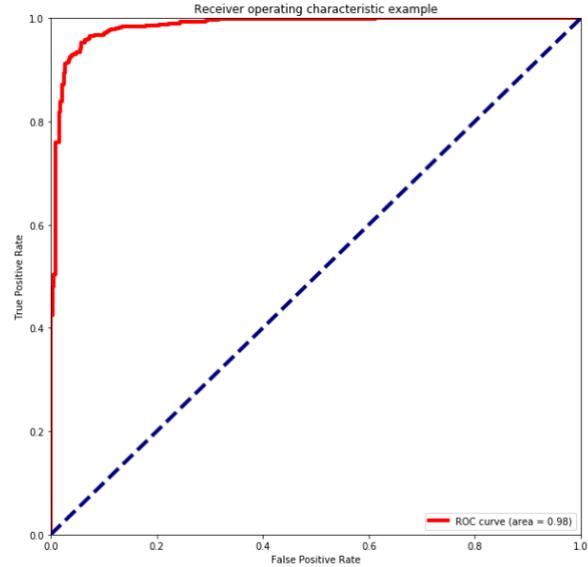

(b) Logistic Regression

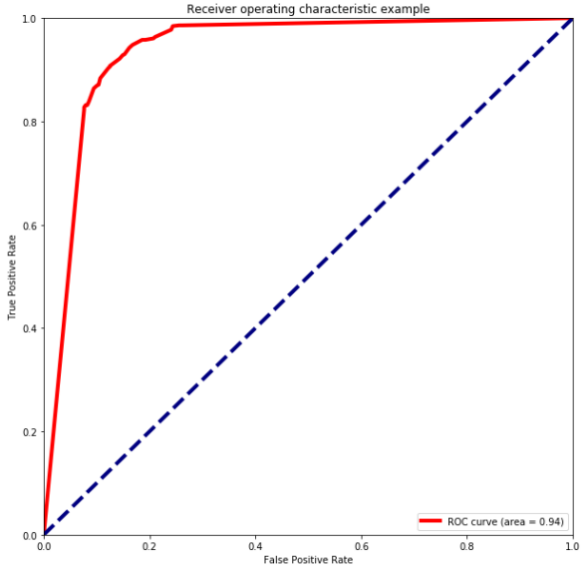

(c) Decision Tree

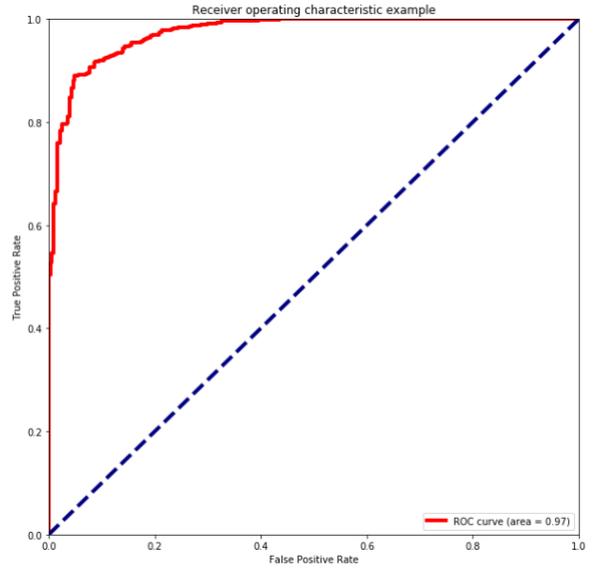

(d) SVM

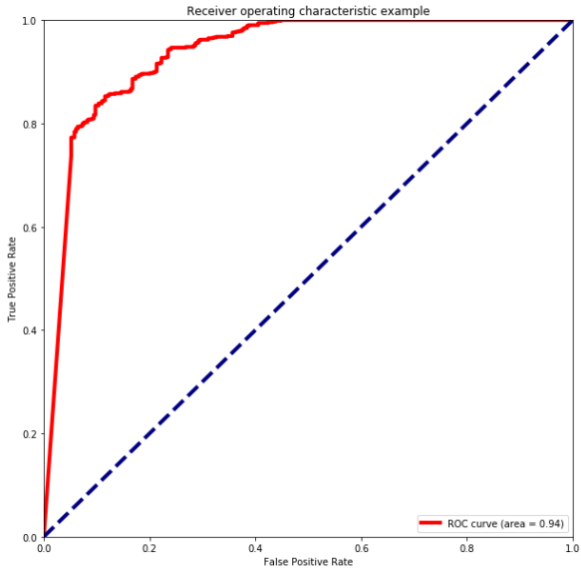

(e) KNN

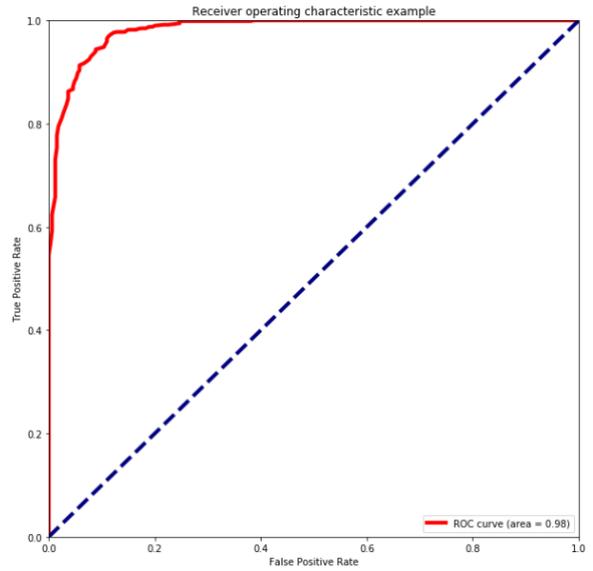

(f) Random Forest

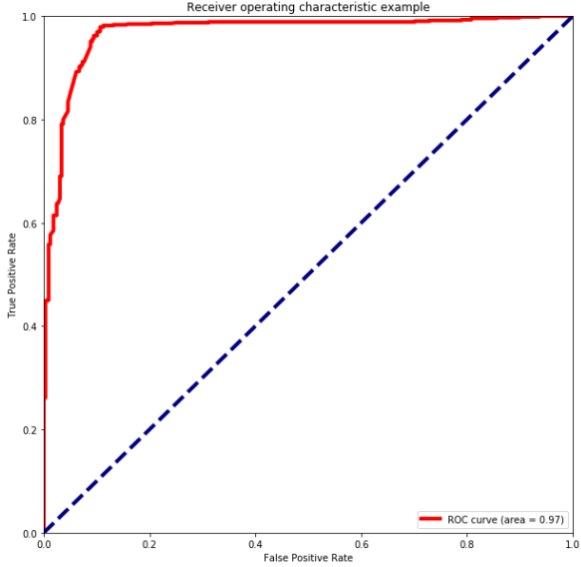
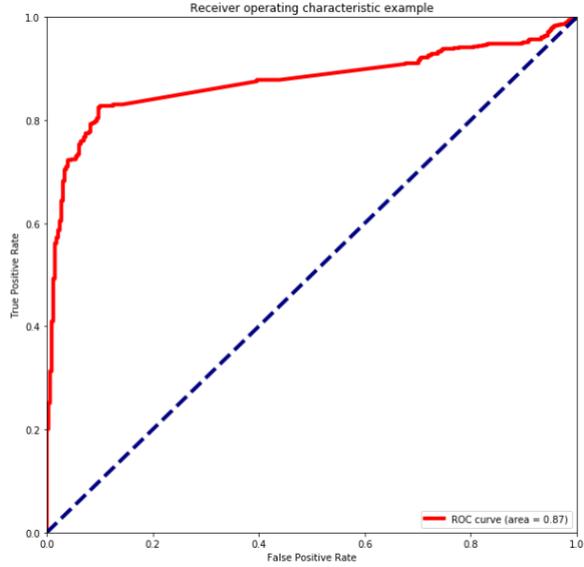

(g) AdaBoost                                                                 (h) MNN

**Fig. 13** Examples of receiver operating characteristic (ROC) graphs

**Table 7** ROC-AUC value of the classification

|  | Naïve Bayes | Logistic Regression | Decision Tree | SVM |
|---|---|---|---|---|
| Tornado | 0.91 | 0.79 | 0.64 | 0.79 |
| Hurricane Sandy | 0.96 | 0.91 | 0.82 | 0.86 |
| Floods | 0.94 | 0.74 | 0.64 | 0.76 |
| Blizzard | 0.90 | 0.79 | 0.64 | 0.79 |
| Matthew | 0.92 | 0.98 | 0.94 | 0.97 |
| Hurricane Harvey | 0.87 | 0.94 | 0.80 | 0.93 |
| Hurricane Michael | 0.96 | 0.96 | 0.80 | 0.95 |
| Wildfires | 0.95 | 0.85 | 0.73 | 0.86 |
| Hurricane Dorian | 0.87 | 0.97 | 0.93 | 0.97 |
| Average | 0.92 | 0.88 | 0.77 | 0.88 |
|  | KNN | Random Forest | AdaBoost | MNN |
| Tornado | 0.70 | 0.80 | 0.72 | 0.67 |
| Hurricane Sandy | 0.80 | 0.84 | 0.86 | 0.87 |
| Floods | 0.70 | 0.78 | 0.71 | 0.70 |
| Blizzard | 0.72 | 0.75 | 0.74 | 0.69 |
| Matthew | 0.94 | 0.98 | 0.97 | 0.87 |
| Hurricane Harvey | 0.77 | 0.93 | 0.71 | 0.92 |
| Hurricane Michael | 0.86 | 0.94 | 0.96 | 0.92 |
| Wildfires | 0.70 | 0.87 | 0.87 | 0.79 |
| Hurricane Dorian | 0.92 | 0.98 | 0.95 | 0.95 |
| Average | 0.79 | 0.87 | 0.83 | 0.82 |

Since the public has more negative reviews than positive reviews, and the analysis of negative reviews helps us improve the efficiency of disaster relief, negative reviews are assigned the label of 1. Therefore, the True Positive Rate (TPR) and False Positive Rate (FPR) are calculated based on negative reviews. In our research, a comparison of different ROC-AUC values is made. We find that that the Naïve Bayes model has the best average ROC-AUC value, followed by the Logistics Regression model, Random Forest model, SVM model, AdaBoost model, MNN model,

KNN model, and Decision Tree model. This result is almost the same as before when we compare different prediction accuracies, which validates our previous assumption – the linear and logistics machine learning models have a better classification performance in terms of sentiment analysis. In addition, the analysis results illustrate that input data can affect the performance of classification. For example, when classifying tweets using the Hurricane Matthew and Hurricane Dorian datasets, which have less noise, we can obtain a higher ROC-AUC value than others. Therefore, whether the collected data is clean is also one of the major factors that affect the classification performance of the machine learning model. Finally, we find that machine learning models' performances are closely related to their parameters. For example, when using GridsearchCV and cross-validation to find the optimal settings of the model, it is difficult to consider all the combinations of parameters because some parameters are continuous values (like c-value in logistic regression and n-estimators in random forests). Consequently, if the user wants to evaluate all the combinations, the computational time will tend to infinity. In practice, people always specify an interval for these continuous parameters to avoid this problem. However, they are likely to miss the optimal settings of the model so that we cannot conclude the Naïve Bayes model must perform better than other machine learning models.

### 5.4.3 Discussion of machine learning models

Different machine learning models have other properties. For example, Naïve Bayes models are easy to implement and can scale with the dataset. Logistic Regression models can be regularized to avoid overfitting, SVM models are effective in high dimensional spaces, and so on. In general, the more complex the machine learning model, the higher the classification accuracy, but their disadvantage is that the longer computation time. To improve the efficiency of disaster management, governments and relief agencies need an analytical approach that is both short time-consuming and high prediction accuracy to deal with the uncertainties of natural disasters related to social media information. Only in this way can people provide rational, real-time, and efficient disaster relief for each natural disaster. Therefore, this paper presents a comparison of the average computational timing and prediction accuracy of the eight machine learning models. Computations were carried out on a MacBook Pro desktop with 2.3 GHz Quad-Core Intel Core i5 processor under a macOS Catalina environment. As shown in Figure 14, we find that the KNN and SVM model has a much higher average computational time than other machine learning models, followed by the AdaBoost model, Random Forest model, MNN model, Logistics Regression model, Decision Tree model, and Naïve Bayes model. While complex models take longer time to predict simple machine learning models, their prediction accuracy is not as good as the latter. For example, the average computation time of the KNN model is 1164.94 seconds, but its average prediction accuracy is only 79%. In contrast, the Naïve Bayes model has an average computation time of 9.02 seconds and a high average prediction accuracy of 92%. The findings suggest that not always the more complex the model, the better the results.

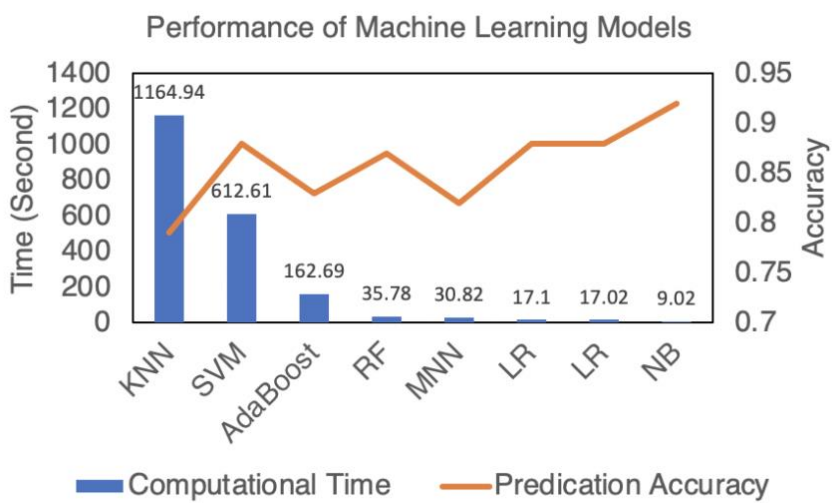

**Fig. 14** Comparison of computational Time

### 6. Conclusion

In this research, a social media data-driven analytics is studied for improving disaster response efficiency, which focuses on investigating public sentimental characteristics via machine learning techniques. The contributions of this research include: 1) exploring and comparing various machine learning models for classifying disaster-related social media data; 2) analyzing essential requirements (e.g., food, housing, transportation, and medical supplies) that victims need during disaster response; 3) conducting a set of real-world instances for understanding changes of public opinions on disaster response, from the perspective of different natural disasters and the most common disaster with disparate time series; and 4) developing a natural disaster dataset with sentiment labels, which contains 41,993 Twitter data about natural disasters in the United States.

We find that the amount of Twitter data collected increases over time, which shows social media is promised to be a critical platform for extracting disaster-related information. Although more than half of the tweets are negative reviews, there have been more and more positive reviews in recent years. Besides, we find that there is a difference between people's essential needs when natural disasters happen. Therefore, relief agencies should provide targeted assistance based on the results of sentiment analysis. Also, depending on whether natural disasters can be accurately predicted, the trend for people to post/obtain relevant information on the Internet is various. To collect clean and comprehensive data, researchers need to clarify the characteristics of natural disasters before conducting an analysis.
Last but not least, according to the comparison of machine learning models, we find that the performance of the model adjusted for the optimal settings is significantly better than that of the existing and lexicon-based machine learning models. Also, we find the Naïve Bayes is the model with the best performance from the perspective of classification accuracy, confusion matrix, and ROC-AUC graph in our research. Moreover, we also notice the differences between models' factors, such as activation function, estimators, and even input data, can cause different results. Therefore, relief agencies or emergency officers should conduct a comparative analysis of machine learning models to get better conclusions when analyzing disaster response problems.

In contrast to previous research, this paper provides a more specific definition of positive and negative sentiments when it comes to sentiment analysis. Also, while researchers pay extensive attention to sentiment analysis, this paper serves better disaster response by mining more useful social media information, such as the proportion of positive and negative tweets in different natural disasters, the proportion of tweets regarding different essential needs, public attitudes towards crucial relief supplies, and the number of tweets in various natural disasters. As time changes, more and more people are willing to use social media platforms to convey their feelings and demands. By studying this information from the public, this paper can provide accurate and timely disaster relief guidance and advice to governments and relief agencies. Besides, this paper also compares different machine learning models and finds that not the more complex the model, the better it works. We have found that some simple models are instead more efficient; they have not only shorter computation times, but also higher prediction accuracy. It is a considerable advantage for prompting the deployment of relief supplies and planning disaster response.

This paper focuses on social media data-driven analytics for improving disaster response efficiency, especially for essential relief supplies, such as housing, transportation, food, and medical supplies. While previous studies (Olanrewaju et al. 2020; Hu et al. 2019; Hu and Dong 2019) have studied supplier selection and pre-positioning strategies in humanitarian relief. We are inspired to expand both us and existing research as an integrated disaster-related social media analytics platform for better disaster management. We will try to combine the advantages of different machine learning models to present a more efficient analysis method.

**Funding** Not applicable.
**Conflicts of interest** Authors declare that they have no conflict of interest.
**Availability of data and material** We are making our dataset available to the research community: https://github.com/Dong-UTIL/Natural-Hazards-Twitter-Dataset
**Code availability** We are making our dataset available to the research community: https://github.com/Dong-UTIL/Natural-Hazards-Twitter-Dataset

**Reference**
Alam F, Ofli F, Imran M, and Aupetit M (2018) A twitter tale of three hurricanes: Harvey, irma, and maria. *arXiv preprint arXiv:1805.05144.*
De Albuquerque JP, Herfort B, Brenning A, and Zipf A (2015) A geographic approach for combining social media and authoritative data towards identifying useful information for disaster management. *International journal of geographical information science*, 29(4), 667-689.
Annett M and Kondrak G (2008) A comparison of sentiment analysis techniques: Polarizing movie blogs. *In Conference of the Canadian Society for Computational Studies of Intelligence (pp. 25-35). Springer, Berlin, Heidelberg.*

Ansari MZ, Aziz MB, Siddiqui MO, Mehra H, and Singh KP (2020) Analysis of political sentiment orientations on twitter. *Procedia Computer Science*, 167, 1821-1828.

Agarwal A, Xie B, Vovsha I, Rambow O, and Passonneau RJ (2011) Sentiment analysis of twitter data. In Proceedings of the workshop on language in social media, pp. 30-38.

Bai H and Yu G (2016) A Weibo-Based Approach to Disaster Informatics: Incidents Monitor in Post-Disaster Situation via Weibo Text Negative Sentiment Analysis. *Natural Hazards* 83(2): 1177–96.

Batrinca B and Treleaven PC (2014) Social Media Analytics: A Survey of Techniques, Tools and Platforms. *AI and Society* 30(1): 89–116.

Beigi, G, Hu X, Maciejewski R, and Liu H (2016) An Overview of Sentiment Analysis in Social Media and Its Applications in Disaster Relief. *Studies in Computational Intelligence* 639: 313–40.

Xiang B and Zhou L (2014) Improving twitter sentiment analysis with topic-based mixture modeling and semi-supervised training. *In Proceedings of the 52nd Annual Meeting of the Association for Computational Linguistics*, Volume 2: Short Papers, pp. 434-439.

Buscaldi D and Hernández-Farias I (2015) Sentiment Analysis on Microblogs for Natural Disasters Management: A Study on the 2014 Genoa Floodings. *WWW 2015 Companion - Proceedings of the 24th International Conference on World Wide Web*: 1185–88.

Calderon NA, Arias-Hernandez R, and Fisher B (2014) Studying Animation for Real-Time Visual Analytics: A Design Study of Social Media Analytics in Emergency Management. *Proceedings of the Annual Hawaii International Conference on System Sciences* (1): 1364–73.

Caragea C, Squicciarini A, Stehle S, Neppalli K, and Tapia A. (2014) Mapping moods: Geo-mapped sentiment analysis during hurricane Sandy. *Proceedings of the international ISCRAM conference*.

Crooks A, Croitoru A, Stefanidis A, and Radzikowski J (2013) Earthquake: Twitter as a distributed sensor system, *T Gis* 17 (1) 124–147.

Chen M, Chen W, and Ku L (2018) Application of Sentiment Analysis to Language Learning. *IEEE Access*. PP. 1-1. 10.1109/ACCESS.2018.2832137.

Earle PS, Bowden DC, Guy M (2012) Twitter earthquake detection: earthquake monitoring in a social world, *Annals of Geophysics, 54(6).* https://doi.org/10.4401/ag- 5364.

Gadekallu T, Soni A, Sarkar D, and Kuruva L (2019) Application of Sentiment Analysis in Movie reviews. *Sentiment Analysis and Knowledge Discovery in Contemporary Business*. pp. 77-90. 10.4018/978-1-5225-4999-4.ch006.

Gao H, Barbier G, and Goolsby R (2011) Harnessing the Crowdsourcing Power of Social Media for Disaster Relief. *IEEE Intelligent Systems* 26(3): 10–14.

Glorot X, Bordes A, and Bengio Y (2011) Domain adaptation for large-scale sentiment classification: A deep learning approach. *Proceedings of the 28th international conference on machine learning (ICML-11)*.

Hamilton WL, Clark K, Leskovec J, and Jurafsky D (2016) Inducing domain-specific sentiment lexicons from unlabeled corpora. *EMNLP 2016 - Conference on Empirical Methods in Natural Language Processing*. 595–605.

Haworth B (2016) Emergency management perspectives on volunteered geographic information: Opportunities, challenges and change. *Computers, Environment and Urban Systems*, 57, 189–198.

Hatzivassiloglou V and Wiebe MJ (2000) Effects of Adjective Orientation and Gradability on Sentence Subjectivity. : 299–305.

Hu S, and Dong Z (2019) Supplier Selection and Pre-Positioning Strategy in Humanitarian Relief. *Omega (United Kingdom)* 83: 287–98. https://doi.org/10.1016/j.omega.2018.10.011.

Hu S, Han C, Dong Z, and Meng L (2019) A Multi-Stage Stochastic Programming Model for Relief Distribution Considering the State of Road Network. *Transportation Research Part B: Methodological* 123: 64–87. https://doi.org/10.1016/j.trb.2019.03.014.

Hughes AL, Denis LA, Palen L, and Anderson MK (2014) Online Public Communications by Police & Fire Services during the 2012 Hurricane Sandy. *Conference on Human Factors in Computing Systems - Proceedings*: 1505–14.

Imran M, Castillo C, Lucas J, Meier P, and Vieweg S (2014) AIDR: Artificial Intelligence for Disaster Response. *WWW 2014 Companion - Proceedings of the 23rd International Conference on World Wide Web*: 159–62.

Kennedy A and Inkpen D (2006) Sentiment Classification of Movie Reviews Using Contextual Valence Shifters. *Computational Intelligence* 22(2): 110–25.

Kiritchenko S, Zhu X, and Mohammad SM (2014) Sentiment analysis of short informal texts. *Journal of Artificial Intelligence Research*, 50.

Kryvasheyeu Y, Chen H, Obradovich N, Moro E, Van Hentenryck P, Fowler J, and Cebrian M (2016) Rapid assessment of disaster damage using social media activity. *Science Advances*, 2(3).

Lee JC, Chung JH, and Kim SJ (2019) The Relationship among Meteorological, Agricultural, and in Situ News-Generated Big Data on Droughts. *Natural Hazards* 98(2): 765–81. https://doi.org/10.1007/s11069-019-03729-7.

Li J, Stephens KK, Zhu Y, and Murthy D (2019) Using Social Media to Call for Help in Hurricane Harvey: Bonding Emotion, Culture, and Community Relationships. *International Journal of Disaster Risk Reduction* 38(June): 101212. https://doi.org/10.1016/j.ijdrr.2019.101212.

Li Z, Wang C, Emrich CT, and Guo D (2018) A novel approach to leveraging social media for rapid flood mapping: a case study of the 2015 South Carolina floods. *Cartography and Geographic Information Science*, 45(2), 97-110.

Lindsay BR (2012) Social Media and Disasters: Current Uses, Future Options, and Policy Considerations. *Social Media and Disasters: Uses, Options, Considerations*: 1–14.

Liu B (2012) Sentiment analysis and opinion mining. *Synthesis Lectures on Human Language Technologies*, 5(1), 1–184.

Liu L, Preotiuc-Pietro D, Samani ZR, Moghaddam ME, and Ungar L (2016) Analyzing Personality through Social Media Profile Picture Choice. *Proceedings of the 10th International Conference on Web and Social Media, ICWSM 2016* (Icwsm): 211–20.

Liu D and Lei L (2018) The appeal to political sentiment: an analysis of donald trump's and hillary clinton's speech themes and discourse strategies in the 2016 us presidential election. *Discourse, Context & Media*, 25.

Martí P, Serrano-Estrda L, and Nolasco-Cirugeda A (2018) Social Media Data: Challenges, Opportunities and Limitations in Urban Studies. *Computers, Environment and Urban Systems*. 74. 10.1016/j.compenvurbsys.2018.11.001.

Mandel B, Culotta A, Boulahanis J, Stark D, Lewis B, and Rodrigue J (2012) A Demographic Analysis of Online Sentiment during Hurricane Irene. *Proceedings of the 2012 Workshop on Language in Social Media* (Lsm): 27–36.

Meng L and Dong Z (2020) Natural Hazards Twitter Dataset. *http://arxiv.org/abs/2004.14456*.

Nagy A and Stamberger J (2012) Crowd sentiment detection during disasters and crises. *Proceedings of the international ISCRAM conference* (pp. 1–9).

Nazer TH, Xue G, Ji Y, and Liu H (2017) Intelligent disaster response via social media analysis-a survey. *ACM SIGKDD Explorations Newsletter*, 19(1), 46–59.

Neppalli K, Caragea C, Squicciarini A, Tapia A, and Stehle S (2017) Sentiment analysis during Hurricane Sandy in emergency response. *International Journal of Disaster Risk Reduction*, 21, 213–222.

Odlum M and Yoon S (2015) What Can We Learn about the Ebola Outbreak from Tweets? *American Journal of Infection Control* 43(6): 563–71. http://dx.doi.org/10.1016/j.ajic.2015.02.023.

Olanrewaju OG, Dong Z, and Hu S (2020) Supplier Selection Decision Making in Disaster Response. *Computers and Industrial Engineering* 143: 106412. https://doi.org/10.1016/j.cie.2020.106412.

Olteanu A, Vieweg S, and Castillo C (2015) What to expect when the unexpected happens: Social media communications across crises. *CSCW 2015 - proceedings of the 2015 ACM international conference on computer-supported cooperative work and social computing* (pp. 994–1009).

Pak A and Paroubek P (2010) Twitter as a Corpus for sentiment analysis and opinion mining. *Proceedings of the seventh international conference on language resources and evaluation* (pp. 1320–1326). Valletta, Malta: European Language Resources Association (ELRA).

Pan S and Yang Q (2010) A survey on transfer learning. *IEEE Transactions on Knowledge and Data Engineering*, 22, 1345–1359.

Pang B and Lee L (2004) A sentimental education: Sentiment analysis using subjectivity summarization based on minimum cuts. *Proceedings of the 42nd annual meeting on Association for Computational Linguistics*.

Pang B and Lee L (2008) 2 *Opinion Mining and Sentiment Analysis: Foundations and Trends in Information Retrieval*.

Power R, Robinson B, Colton J, and Cameron M (2014) Emergency situation awareness: twitter case studies, in: C. Hanachi, F. Bénaben, F. Charoy (Eds.), Information Systems for Crisis Response and Management in Mediterranean Countries. *Springer International Publishing, Cham,* pp. 218–231.

Ranasinghe H and Halgamuge M (2020) Twitter Sentiment Data Analysis of User Behavior on Cryptocurrencies: Bitcoin and Ethereum. *Analyzing Global Social Media Consumption.*10.4018/978-1-7998-4718-2.ch015.

Ragini JR, Anand PMR, and Bhaskar V (2018) Big Data Analytics for Disaster Response and Recovery through Sentiment Analysis. *International Journal of Information Management* 42: 13–24. https://doi.org/10.1016/j.ijinfomgt.2018.05.004.

Shabaz M and Garg U (2020) Clustering Yelp's sentiment data through various approaches and estimating the error rate. *Materials Today: Proceedings*. 10.1016/j.matpr.2020.09.346.

Shibuya Y and Tanaka H (2018) Public Sentiment and Demand for Used Cars after A Large-Scale Disaster: Social Media Sentiment Analysis with Facebook Pages. *arXiv preprint arXiv:1801.07004.*

Song Y, Ji Q, Du YJ, and Geng, JB. (2019). The dynamic dependence of fossil energy, investor sentiment and renewable energy stock markets. *Energy economics*, 84(Oct.), 104564.1-104564.15.

Si S (2015) Social Media and Its Role in Marketing. *Business and Economics Journal* 07(01).

Smith L, Liang Q, James P, and Lin W (2017) Assessing the utility of social media as a data source for flood risk management using a real-time modelling framework, *J. Flood Risk Manag*. 10 (3) 370–380.

Tang D, Wei F, Yang N, Zhou M, Liu T, and Qin B (2014) Learning sentiment-specific word embedding for twitter sentiment classification. *52nd annual meeting of the Association for Computational Linguistics, ACL 2014 - proceedings of the conference*. Vol. 1. pp. 1555–1565.

To H, Agrawal S, Kim SH, and Shahabi C (2017) On identifying disaster-related tweets: Matching-based or learning-based. *Proceedings - 2017 IEEE 3rd international conference on multimedia big data*, BigMM 2017 pp. 330–337.

Turner-McGrievy G, Karami A, Monroe C, and Brandt HM (2020) Dietary Pattern Recognition on Twitter: A Case Example of before, during, and after Four Natural Disasters. *Natural Hazards*. https://doi.org/10.1007/s11069-020-04024-6.

Vuong NB and Suzuki Y (2020). Impact of financial development on sentiment-return relationship: insight from asia-pacific markets 1. *Borsa Istanbul Review*, 20( 2), 95-107.

Wang B and Zhuang J (2018) Rumor Response, Debunking Response, and Decision Makings of Misinformed Twitter Users during Disasters. *Natural Hazards* 93(3): 1145–62. https://doi.org/10.1007/s11069-018-3344-6.

Wang Y, Taylor JE, and Garvin MJ (2020) Measuring Resilience of Human-Spatial Systems to Disasters: Framework Combining Spatial-Network Analysis and Fisher Information. *Journal of Management in Engineering* 36(4).

Wang Y and Taylor JE (2018) Coupling sentiment and human mobility in natural disasters: A Twitter-based study of the 2014 South Napa Earthquake. *Natural Hazards*, 92(2), 907–925. https://doi.org/10.1007/s11069-018-3231-1.

Wu F and Huang Y, Yuan Z (2017) Domain-specific sentiment classification via fusing sentiment knowledge from multiple sources. *Information Fusion*, 35, 26–37.

Yin J, Lampert A, Cameron M, Robinson B, and Power R (2012) Using social media to enhance emergency situation awareness. *IEEE intelligent systems*, (6), 52-59.

Zhao J and Gui X (2017) Comparison research on text pre-processing methods on twitter sentiment analysis. *IEEE Access*, PP(99), 1-1.